\documentclass[epj,nopacs,fleqn]{svjour}
\journalname{Eur. Phys. J. C}
\smartqed

\usepackage[T1]{fontenc}
\usepackage{cite,epsfig,amsmath,graphicx,amssymb,listings,slashed}
\usepackage[colorlinks,citecolor=blue,urlcolor=blue,linkcolor=blue]{hyperref}
\usepackage[frozencache]{minted}
\usepackage{xpatch}
\xpatchcmd{\mintinline}{\begingroup}{\begingroup\let\itshape\relax}{}{}
\xpatchcmd{\minted}{\VerbatimEnvironment}{\VerbatimEnvironment\let\itshape\relax}{}{}
\usepackage{tcolorbox}
\usepackage{etoolbox}
\BeforeBeginEnvironment{lstlisting}{
\begin{tcolorbox}[left=2mm, right=0mm, top=0mm, bottom=0mm]\small}
\AfterEndEnvironment{lstlisting}{\end{tcolorbox}}

\usepackage{listings}
\usepackage{hhline}
\usepackage{makecell, booktabs}
\usepackage{stfloats}

\begin{document}

\title{{\sc HEP ML Lab}: An end-to-end framework for applying machine learning into phenomenology studies}
\author{Jing Li\inst{1} \and Hao Sun\inst{1}}
\institute{Department of Physics, Dalian University of Technology, Dalian, 116024, China}
\date{Received: date / Revised version: date}

\abstract{
Recent years have seen the development and growth of machine learning in high
energy physics. There will be more effort to continue exploring its full
potential. To make it easier for researchers to apply existing algorithms and
neural networks and to advance the reproducibility of the analysis, we develop
the {\sc HEP ML Lab} (\texttt{hml}), a Python-based, end-to-end framework for
phenomenology studies. It covers the complete workflow from event generation to
performance evaluation, and provides a consistent style of use for different
approaches. We propose an observable naming convention to streamline the data
extraction and conversion processes. In the {\sc Keras} style, we provide the
traditional cut-and-count and boosted decision trees together with neural
networks. We take the $W^+$ tagging as an example and evaluate all built-in
approaches with the metrics of significance and background rejection. With its
modular design, {\sc HEP ML Lab} is easy to extend and customize, and can be
used as a tool for both beginners and experienced researchers.
}

\titlerunning{{\sc HEP ML Lab}}
\authorrunning{Jing Li and Hao Sun}
\maketitle

% Introduction --------------------------------------------------------------- %
\section{Introduction} \label{introduction}

In recent years, with the continuous accumulation of data from the Large Hadron
Collider experiments, the search for new physics has posed higher demands.
Machine learning techniques, due to their outstanding capabilities in data
analysis and pattern recognition, have received wide-spread attention,
exploration, and application in high-energy physics, such as jet tagging tasks
\cite{Cogan:2014oua,Almeida:2015jua,deOliveira:2015xxd,Baldi:2016fzo,Komiske:2016rsd,Kasieczka:2017nvn,Dery:2017fap,Louppe:2017ipp,Butter:2017cot,Metodiev:2017vrx,Aguilar-Saavedra:2017rzt,Moore:2018lsr,Heimel:2018mkt,Komiske:2018cqr,Qu:2019gqs,Kasieczka:2019dbj,Moreno:2019bmu,Chen:2019uar,Mikuni:2020wpr,Lee:2020qil,Dreyer:2020brq,Anzalone:2022hrt,Choi:2023slq},
rapid generation of simulated events
\cite{deOliveira:2017pjk,Paganini:2017hrr,Paganini:2017dwg,Baldi:2020hjm,Jiang:2024ohg,Kobylianskii:2024ijw}.
More applications can refer to this review \cite{Feickert:2021ajf}.

Typically, the process of research involving machine learning models in
high-energy physics comprises four steps: data generation, dataset construction,
model training, and performance evaluation. In this process, cooperation between
various software is often required. For instance, use {\sc MadGraph5\_aMC}
\cite{Alwall:2014hca} for generating simulated events, {\sc Pythia8}
\cite{Sjostrand:2014zea} for simulating parton showering, {\sc Delphes}
\cite{deFavereau:2013fsa} for fast simulating detector effects, {\sc ROOT}
\cite{rene_brun_2020_3895860} for data processing, and subsequently building
neural networks with deep learning frameworks such as {\sc PyTorch}
\cite{Ansel_PyTorch_2_Faster_2024} and {\sc TensorFlow}
\cite{tensorflow2015-whitepaper}. For researchers new to high-energy physics,
learning and using these software tools pose a significant challenge, while for
experienced researchers, switching between different software can be a tedious
task. Such a process inevitably increases the complexity of computational
results, making them potentially difficult to replicate, leading to difficulties
in result comparison in subsequent research.

Currently, some efforts have been made to simplify the entire process:
{\sc pd4ml} \cite{Benato:2021olt} includes five datasets: Top Tagging
Landscape, Smart Background, Spinodal or Not, EoS, Air Showers, and provides a
set of concise application programming interfaces (API) for importing them;
{\sc MLAnalysis} \cite{Guo:2023nfu} can convert LHE and LHCO files generated
by {\sc MadGraph5\_aMC} into datasets, and has three built-in machine
learning algorithms: isolation forest (IF), nested isolation forest (NIF), and
k-means anomaly detection (KMAD); {\sc MadMiner} \cite{Brehmer:2019xox}
offers a complete process for inference tasks \cite{Brehmer:2019bvj}, and
internally encapsulates the necessary simulation software, as well as neural
networks based on {\sc PyTorch}. These frameworks significantly reduce the
workload related to specific tasks but still have areas that could be improved.

\begin{figure*}[h]
\centering
\includegraphics[width=0.8\textwidth]{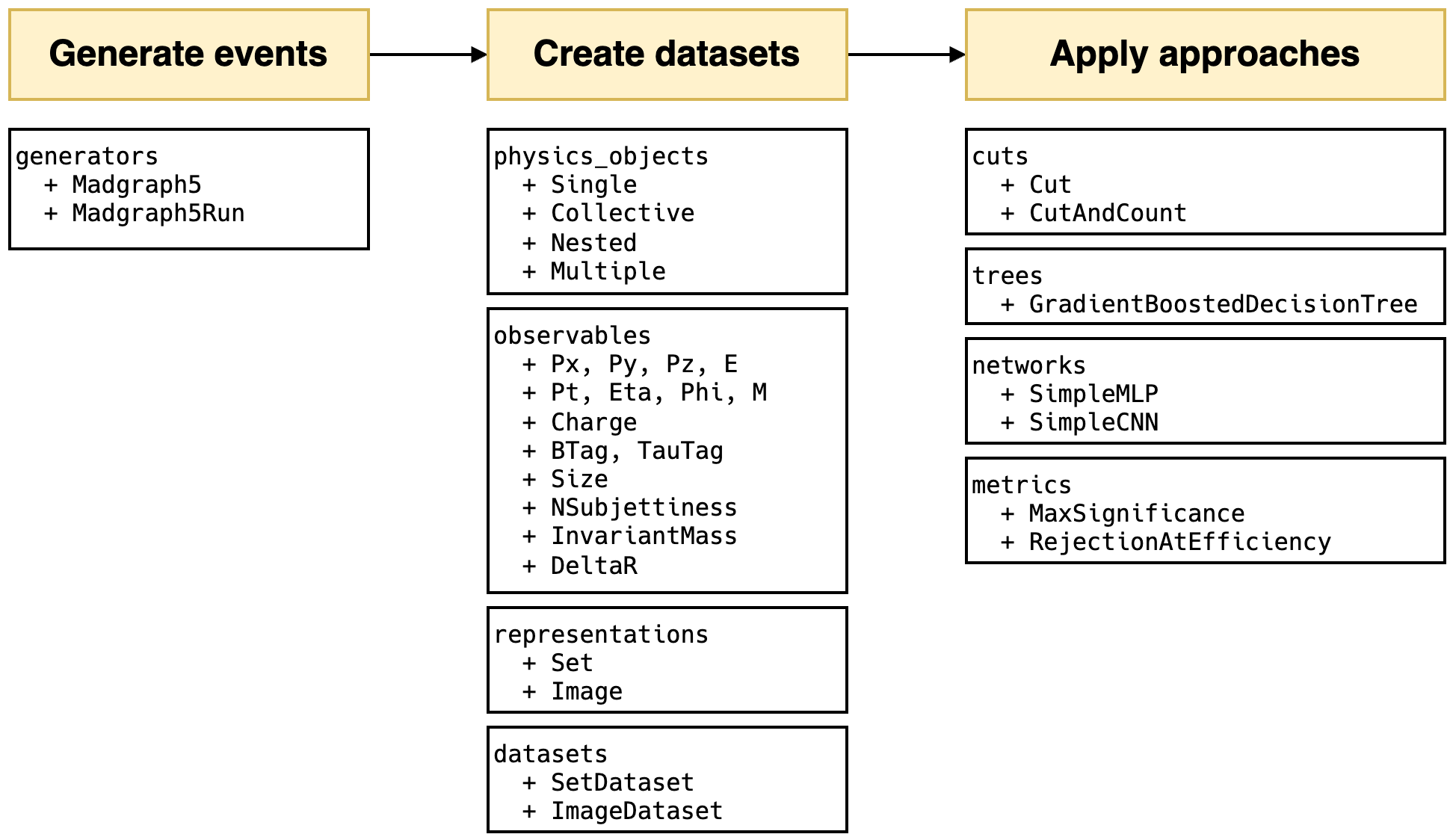}
\caption{All modules in the \texttt{hml} framework and main classes in each module.}
\label{fig:hml_modules}
\end{figure*}

{\sc HEP ML Lab}, developed in Python, encompasses an end-to-end complete
process. All modules are shown briefly in figure \ref{fig:hml_modules}. {\sc
MadGraph5\_aMC} is minimally encapsulated for event generation, such as defining
processes, generating Feynman diagrams, and launching runs. In the transition
from events to datasets, we introduced an \textit{observable naming convention}
that directly links physical objects with observables, facilitating users to
directly use the names of observables to retrieve corresponding values. This
convention can further apply to the definitions of cuts. Inspired by the
expression form of cuts in {\sc uproot} \cite{jim_pivarski_2020_3952728}, we
expand the corresponding syntax to support filtering at the event level, using
veto to define events that need to be removed and more complex custom
observables. When creating datasets with different representations, this naming
rule still applies. In the current version, users can easily create set and
image datasets, and for images, we also offer a rich set of functions for
preprocessing and displaying.

In the part of machine learning, we introduce two basic deep learning models:
simple multi-layer perceptron and simple convolutional neural network. Both have
fewer than ten thousand parameters, providing a baseline for classification
performance. These models are implemented using {\sc Keras}
\cite{chollet2015keras} without any custom modifications, making it easy to
expand to other existing models. Additionally, we integrate two traditional
approaches, cut-and-count, and gradient boosted decision tree, ensuring
compatibility with {\sc Keras}. After different approaches are trained, we
provide physics-based evaluation metrics: signal significance and background
rejection rate at fixed signal efficiency, to assess their performance.

The structure of the paper is as follows. The section \ref{generate_events}
introduces the wrapper class of {\sc MadGraph5\_aMC} to generate events. In
the section \ref{create_datasets}, we describe the observable naming convention
and show step by step how it is used to extract data from events and extended to
filter data and to create datasets. Three types of approaches: cut and count,
decision trees, and neural networks available now are shown in the section
\ref{apply_approaches}. Physics-inspired metrics are also included there. In the
section \ref{tagging}, we demonstrate the effectiveness of the framework by a
simple and complete W boson tagging as a case study. Finally, we conclude the
paper and discuss the future work in the section \ref{summary}.

% Generate events ------------------------------------------------------------ %
\section{Generate events} \label{generate_events}

All phenomenological studies generally start from simulating collision events,
for example using {\sc MadGraph5\_aMC}. The \texttt{generators} module
provides a wrapper for parts of its core functionalities, aiming to facilitate
its integration into Python scripts for customized setting requirements.

\begin{listing}[h]
\begin{lstlisting}
from hml.generators import Madgraph5

g = Madgraph5(executive="mg5_aMC", verbose=1)
\end{lstlisting}
\caption{Initialize \texttt{Madgraph5}.}
\label{code:init_mg5}
\end{listing}

In code example \ref{code:init_mg5}, users need to pass the executable path to
the \texttt{executive} parameter to ensure commands can be sent  to it. The
\texttt{verbose} parameter controls whether to display intermediate outputs,
with the default value of 1, meaning they are displayed, consistent with the
output seen when using {\sc MadGraph5\_aMC} in the terminal. After
initialization, we can use its various methods to simulate commands entered in
the terminal as shown in code example \ref{code:generate_processes}.

\begin{listing}[h]
\begin{lstlisting}
g.import_model(model)
g.define(expression)
g.generate(*processes)
g.display_diagrams(diagram_dir="Diagrams", overwrite=True)
g.output(output_dir, overwrite=True)
\end{lstlisting}
\caption{Methods of \texttt{Madgraph5} to generate processes.}
\label{code:generate_processes}
\end{listing}

During the process generation, we first need to use the \texttt{import\_model}
method to import the model file. This method supports passing the path of the
model or the name of the model ({\sc MadGraph5\_aMC} will search for the
model in the models folder or download the model based on its name). Next, use
the \texttt{define} method to define multi-particles, for example,
\texttt{define("j = j b b\textasciitilde")}. Then, in the \texttt{generate}
method, pass in all the processes to be generated without having to input
\texttt{add process} like in the terminal. Here, the asterisk represents the
unpacking operation in Python, and you can directly enter multiple processes
separated by commas \texttt{g.generate("p p > w+ z", "p p > w- z")} without
needing to construct a list with square brackets. Usually, to confirm processes
have been generated as expected, we need to view the Feynman diagrams, at which
point the \texttt{display\_diagrams} method can be used. It saves the generated
Feynman diagrams to the \texttt{diagram\_dir} folder and has already converted
the default eps files into pdf format for convenience. Finally, use the
\texttt{output} method to export the processes to a folder.

\begin{listing}[h]
\begin{lstlisting}
g.launch(
    shower="off",
    detector="off",
    madspin="off",
    settings={},
    decays=[],
    cards=[],
    multi_run=1,
    seed=None,
    dry=False,
)
\end{lstlisting}
\caption{Use \texttt{launch} method and set up all possible parameters for generating events.}
\label{code:launch_runs}
\end{listing}

With the process folder ready, we can start to produce runs to generate
simulated events as shown in code example \ref{code:launch_runs}. The
\texttt{launch} method includes parameters you may need to configure for the
run, where \texttt{shower}, \texttt{detector}, \texttt{madspin} represent
switches for {\sc Pythia8}, {\sc Delphes}, and {\sc Madspin},
respectively, consistent with the options in the terminal's
prompt.\texttt{settings} includes parameters configured in the run card, for
example, \texttt{settings=\{"nevents": 1000, "iseed": 42\}}. While
\texttt{iseed} is the random seed used by {\sc MadGraph5\_aMC} to control the
randomness of the sub-level events, it does not affect {\sc Pythia8} and
{\sc Delphes}. You can specify the \texttt{seed} parameter to uniformly
configure these three, ensuring the cross section, error, and events are fully
reproducible. The \texttt{decays} method is used to set the decay of particles,
for example, \texttt{decays=["w+ > j j", "z > vl vl\textasciitilde"]}. The
\texttt{cards} parameter accepts the path to your pre-configured parameter
files, for example, \texttt{cards=["delphes\_card.dat", "pythia8\_card.dat"]}.
When a large number of events need to be generated, you can set the
\texttt{multi\_run} parameter to create multiple sub-runs for a single run, for
example, setting \texttt{multi\_run=2}, the final event files will be named in
the form of \texttt{run\_01\_1}, \texttt{run\_01\_2}, which is controled by
\texttt{MadEvent}. Note, since {\sc MadGraph5\_aMC} does not recommend generating
more than one million events in a single run, the \texttt{nevents} parameter
in \texttt{settings} should also be set appropriately, as the total number of
events is the result of \texttt{nevents} multiplied by \texttt{multi\_run}.
\texttt{hml} will generate the corresponding valid commands based on your
settings and send them to {\sc MadGraph5\_aMC} running in the background. If
you want to check the actual commands before the run starts, you can set
\texttt{dry=True}, which returns the generated commands instead of starting the
run.

\begin{figure}[h]
  \centering
  \includegraphics[width=0.45\textwidth]{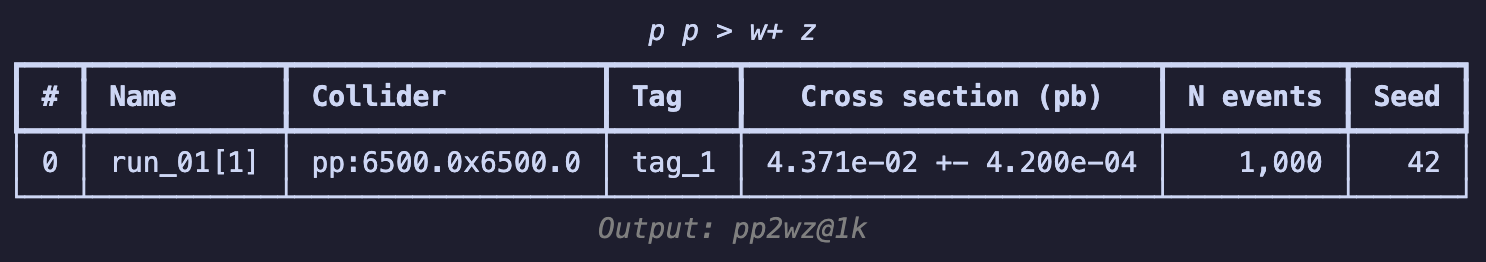}
  \caption{The output of \texttt{summary} method.}
  \label{fig:madgraph5_summary}
\end{figure}

\begin{listing}[h]
\begin{lstlisting}
run = g.runs[0]
# run.name
# len(run.sub_runs)
# run.collider
# run.tag
# run.cross
# run.error
# run.n_events
# run.seed
\end{lstlisting}
\caption{All the information seen in the table can be accessed.}
\label{code:access_run}
\end{listing}

After generating the events, you can use the \texttt{summary} method, i.e.,
\texttt{g.summary()} to print the results in a table as shown in figure
\ref{fig:madgraph5_summary}. The table includes the name of each run, the number
of sub-runs in brackets, colliding particle beam information, tags,
cross-section, error, total number of events, and the random seed. The header
displays process information, and the footnote shows the output's relative path,
essentially consistent with the results you see on the web page.

\begin{listing}[h]
\begin{lstlisting}
for mass in [100, 200, 300]:
    g.launch(
        settings={
            "nevents": 1000,
            "run_tag": f"mass={mass}"},
            "mnh2": mass,
            "wnh2": "auto",
        seed=42,
    )
g.summary()
\end{lstlisting}
\caption{Use a loop to scan the mass of a particle called "nh2" and show the summary.}
\label{code:mass_scan}
\end{listing}

If you wish to continue experimenting with different parameter combinations, you
can use the \texttt{launch} method again, or employ Python's loop statements to
generate a series of combinations to observe the differences in cross-section
under various conditions. When doing so, it is recommended to set short label
names to facilitate subsequent search and analysis like code example
\ref{code:mass_scan} does.

If there're already output files and you can use \texttt{hml} to extract some
information for subsequent use. The class method \texttt{Madgraph5.from\_output}
and the \texttt{Madgraph5Run} will be of great assistance as shown in
\ref{code:from_output}. The former accepts the path to the output folder, which
is the path you enter in the \texttt{output} command of \texttt{MadGraph5}, as
well as the path to the executable file. The latter requires the output folder
path and the name of the run to access information such as cross section and
error. The \texttt{events} method allows for retrieving the paths to all event
files under a run, including sub-runs. Currently, it only supports files in root
format. You can use \texttt{uproot} to open these files for subsequent
processing.

\begin{listing}[h]
\begin{lstlisting}
import uproot
from hml.generators import Madgraph5, Madgraph5Run

g = Madgraph5.from_output(output_dir, executive="mg5_aMC")
run = g.runs[0]

# Or
run = Madgraph5Run(output_dir, name)
print(run)
# Madgraph5Run run_01 (1 sub runs):
#  - collider: pp:6500.0x6500.0
#  - tag: 1k
#  - seed: 42
#  - cross: 0.04371
#  - error: 0.00042
#  - n_events: 1000,

# Open the first root file it finds
uproot.open(run.events(file_format="root")[0])
\end{lstlisting}
\caption{Use \texttt{Madgraph5.from\_output} and \texttt{Madgraph5Run} to access the information.}
\label{code:from_output}
\end{listing}

% Create datasets ------------------------------------------------------------ %
\section{Create datasets} \label{create_datasets}

The leading fat jet's mass, the angular distance between the primary and
secondary jets, the total transverse momentum of all jets, the number of
electrons, etc., all demonstrate that observables are always connected to
certain physical objects. Thus, we propose the observable naming convention: the
name of an observable is a combination of the physical object's name and the
type of observable, connected by a dot, denoted as \texttt{<physics
object>.<observable type>}. In this section, starting from physical objects, we
gradually refine this representation, eventually extending it to the acquisition
of observables, the construction of data representations, and the definitions of
cuts.

\subsection{Physics objects} \label{physics_objects}

Physical objects in {\sc Delphes} are stored in different branches, representing a category rather than a specific instance. Considering that the calculation of many observables involves different types and numbers of physical objects, often utilizing their fundamental four-momentum information, we have categorized physical objects into four types based on their quantity and category:

\begin{enumerate}
    \item \texttt{Single} physical objects, which precisely refer to a specific physical object. For example:
        \begin{itemize}
            \item \texttt{"jet0"} is the leading jet.
            \item \texttt{"electron1"} is the secondary electron.
        \end{itemize}
    \item \texttt{Collective} physical objects, representing a category of physical objects. For example:
        \begin{itemize}
            \item \texttt{"jet"} or \texttt{"jet:"} represents all jets.
            \item \texttt{"electron:2"} represents the first two electrons.
        \end{itemize}
    \item \texttt{Nested} physical objects, formed by free combinations of single and collective physical objects. For example:
        \begin{itemize}
            \item \texttt{"jet.constituents"} represents all constituents of all jets.
            \item \texttt{"jet0.constituents:100"} represents the first 100 constituents of the leading jet.
        \end{itemize}
    \item \texttt{Multiple} physical objects, composed of the previous three types and separated by commas. For example:
        \begin{itemize}
            \item \texttt{"jet0,jet1"} represents the leading and secondary jets.
        \end{itemize}
\end{enumerate}

This naming convention is inspired by the syntax of Python lists. To minimize
the input cost for the user, we discard the original requirement to use square
brackets for receiving indexes or slices: for single physical objects, the type
name is directly connected to the index value; for collective physical objects,
a colon is used to separate the start index from the end index, and the type
name alone represents the whole. Use \texttt{parse\_physics\_object} method to
get the branch and the required index values based on the name of the physical
objects as shown in code example \ref{code:parse_objects}. This design makes
users focus on the physical objects themselves, rather than on how the
corresponding classes should be initialized. In table \ref{tab:physics_objects},
we also summarize all types of physical objects, their initialization
parameters, and examples.

\begin{listing}[h]
\begin{lstlisting}
from hml.physics_objects import parse_physics_object

# Single
obj = parse_physics_object("jet0")
# obj.branch: "jet"
# obj.slices: [slice(0, 1)]

# Collective
obj = parse_physics_object("jet:10")
# obj.branch: "jet"
# obj.slices: [slice(None, 10)]

# Nested
obj = parse_physics_object("jet0.constituents:10")
# obj.branch: "jet.constituents"
# obj.slices: [0, slice(None, 10)]

# Multiple
obj = parse_physics_object("jet0,jet1")
# obj.branch: ["jet0", "jet1"]
# obj.slices: [slice(0, 1), slice(1, 2)]
\end{lstlisting}
\caption{Use \texttt{parse\_physics\_object} method to get the branch and slices of physics objects.}
\label{code:parse_objects}
\end{listing}

\begin{table*}[ht]
\centering
\begin{tabular}{lll}
\toprule
    \textbf{Type} & \textbf{Initialization parameters} & \textbf{Name examples} \\
    \midrule
    \texttt{Single} &  \makecell[l]{\texttt{branch: str,} \\ \texttt{index: int}} & \texttt{"jet0", "muon0"} \\
    \midrule
    \texttt{Collective} & \makecell[l]{\texttt{branch: str,} \\ \texttt{start: int|None}} & \texttt{"jet", "jet1:", "jet:3", "jet1:3} \\
    \midrule
    \texttt{Nested} & \makecell[l]{\texttt{main: str|PhysicsObject,} \\ \texttt{sub: str|PhysicsObject}} &  \texttt{"jet.constituents", "jet0.constituents:100} \\
    \midrule
    \texttt{Multiple} & \makecell[l]{\texttt{all: list[str|Physicsobject]}} & \texttt{"jet0,jet1", "jet0,jet"}\\
\bottomrule
\end{tabular}
\caption{All types of physics objects and their examples.}
\label{tab:physics_objects}
\end{table*}

In this version, physical objects are merely tools for parsing user input and do
not contain any information about observables. Unlike other software packages,
we strictly separate the acquisition of observables from the physical objects.
Physical objects only store information about the connection between observables
and their data sources, not the data itself.

\subsection{Observable} \label{observable}

After defining physical objects, the task of observables is to extract
information from them. In code example \ref{code:parse_objects}, we store all
useful information from a physical object in \texttt{branch} and
\texttt{slices}: the former refers to the corresponding branch name, and the
latter means a specific parts of array-like data. The advantage of doing so is
that when encountering certain physical objects, such as the hundredth jet,
which does not exist, it returns a list of length zero instead of an error. An
empty list will automatically be judged as \texttt{False} when applying cuts,
thereby being skipped.

In table \ref{tab:observables}, we list all the observables currently available.
To avoid remembering exact name of an observable, its name is case-insensitive
and common aliases are added. For example, \texttt{Mass} can be written as
\texttt{mass} or \texttt{m}, and \texttt{NSubjettinessRatio} has the alias
\texttt{tau\{m\}\{n\}}, where the values of \texttt{m} and \texttt{n} are passed
as parameters into the corresponding class. For the transverse momentum,
considering the style in different softwares, we more aliases for its symbol
representation. Moreover, different observables support different types of
physical objects. For example, the \texttt{Size} observable supports collective
physical objects, while the \texttt{AngularDistance} observable supports all
combinations of multi-body objects.

\begin{table*}[h]
  \centering
  \begin{tabular}{llcccc}
  \toprule
      \textbf{Type} & \textbf{Alias} & \textbf{Single} & \textbf{Collective} & \textbf{Nested} & \textbf{Multiple} \\
      \midrule
      \texttt{MomentumX, Px} & \texttt{momentum\_x, px} & \checkmark & \checkmark & \checkmark & \\
      \texttt{MomentumY, Py} & \texttt{momentum\_y, py} & \checkmark & \checkmark & \checkmark & \\
      \texttt{MomentumZ, Pz} & \texttt{momentum\_z, pz} & \checkmark & \checkmark & \checkmark & \\
      \texttt{Energy, E} & \texttt{energy, e} & \checkmark & \checkmark & \checkmark & \\
      \midrule
      \texttt{TransverseMomentum, Pt} & \texttt{transverse\_momentum, pt, pT, PT} & \checkmark & \checkmark & \checkmark & \\
      \texttt{PseudoRapidity, Eta} & \texttt{pseudo\_rapadity, eta} & \checkmark & \checkmark & \checkmark & \\
      \texttt{AzimuthalAngle, Phi} & \texttt{azimuthal\_angle, phi} & \checkmark & \checkmark & \checkmark & \\
      \texttt{Mass, M} & \texttt{mass, m} & \checkmark & \checkmark & \checkmark & \\
      \midrule
      \texttt{Charge} & \texttt{charge} & \checkmark & \checkmark & \checkmark & \\
      \texttt{BTag} & \texttt{b\_tag} & \checkmark & \checkmark & & \\
      \texttt{TauTag} & \texttt{tau\_tag} & \checkmark & \checkmark & & \\
      \texttt{NSubjettiness, TauN} & \texttt{n\_subjettiness,tau\_\{n\}, tau\{n\}} & \checkmark & \checkmark & & \\
      \texttt{NSubjettinessRatio, TauMN} & \texttt{n\_subjettiness\_ratio, tau\_\{m\}\{n\}, tau\{m\}\{n\}} & \checkmark & \checkmark & & \\
      \midrule
      \texttt{Size} & \texttt{size} & & \checkmark & & \\
      \texttt{InvariantMass} & \texttt{invariant\_mass, inv\_mass, inv\_m} & \checkmark & & & \checkmark \\
      \texttt{AngularDistance, DeltaR} & \texttt{angular\_distance, delta\_r} & \checkmark & \checkmark & \checkmark & \checkmark \\
  \bottomrule
  \end{tabular}
  \caption{All types of observables and their supported types of physical objects.}
  \label{tab:observables}
\end{table*}

In code example \ref{code:parse_observables}, we show how to use such an
observable. First initialize the corresponding observables using the
\texttt{parse\_observable} function from the \texttt{observables} module, then
use the \texttt{read} method to extract the values from an event. As the
\texttt{read} returns the object itself, you can take the advantage of method
chaining to define an observable directly followed by reading an event. We also
add extra information when you print the observable itself: its name, shape, and
data type. Internally, \texttt{awkward}\cite{Pivarski_Awkward_Array_2018} is
used for manipulating variable-lengthed jagged arrays. The question mark in the
data type indicates there are missing values (\texttt{None}). The \texttt{var}
appearing in the shape indicates inconsistent lengths, for example, each event
has a varying number of jets and each jet has a varying number of constituents.

\begin{listing}[h]
\begin{lstlisting}
from hml.observables import parse_observable

obs = parse("jet0.pt")
obs.read(events)
print(obs)
# jet.pt: 100 * 1 * ?float32

obs = parse("jet.pt").read(events))
print(obs)
# jet.pt: 100 * var * float32

obs = parse("jet.constituents.pt")
obs.read(events)
# obs.name: "jet.constituents.pt"
# obs.shape: (100, var, var)
# obs.dtype: float32

obs = parse("jet:10.constituents:100.pt")
# obs.shape: (100, 10, 100)
\end{lstlisting}
\caption{Use \texttt{parse\_observable} and \texttt{read} to get the value of observables. \texttt{events} are opened by \texttt{uproot}.}
\label{code:parse_observables}
\end{listing}

While the first dimension of the observable value always represents the number
of events, the shape is generally determined by the related physical objects.
For example, the shape of transverse momentum and other kinematic variables is
exactly as its physics obsject. However, this also depends on how to compute the
observable. For instance, the shape of the \texttt{Size} observable is the
number of physics objects and the second dimension is always $1$, whereas the
shape of \texttt{AngularDistance} depends on the type of physical objects: if
calculating the distance between all jets and the leading fat jet, we will get
an array of shape \texttt{(n\_events, var, 1)}, where \texttt{n\_events}
represents the number of events, \texttt{var} represents a variable number of
jets, and $1$ represents the leading fat jet; if calculating the distance
between the first ten fat jets and all constituents of all jets, we obtain an
array of shape \texttt{(n\_events, 10, var)}. The \texttt{var} now comes from
the number of constituents and the number of jets, which we compress two
dimensions into one. For events that do not have enough physics objects, the
missing values are filled with \texttt{None}.

The built-in observables are only some of the most basic ones, so they may not
be sufficient for every use case. Therefore, we show three examples of building
your own observables. In the first example \ref{code:MET}, when the needed
observable is already stored under a certain branch, you only need to declare
the name of this observable as a class that inherits from \texttt{Observable}.
\texttt{hml} will search for the branch based on the physical object name and
extract the corresponding value based on the slices. Here, we take the
observable \texttt{MET} as an example, which is originally stored under the
branch \texttt{MissingET}. To use the \texttt{parse\_observable} function,
you can call \texttt{register\_observable} function to register an alias for it.
Please note that this implementation requires a physical object, which means
that only by entering \texttt{"missinget0.met"} or \texttt{"MissingET0.MET"} can
the \texttt{parse\_observables} works as normal. Only \texttt{"MET"} or
\texttt{"met"} without an physics object is not allowed. As each event has only
one missing energy physical object, \texttt{"MissingET"} is followed by $0$.

\begin{listing}[h]
\begin{lstlisting}
from hml.observables import (
    Observable,
    parse_observable,
    register_observable,
)

class MET(Observable): ...

register(MET, "MET", "met")

# Both lower and upper cases are acceptable
obs = parse_observable("missinget0.met")
obs = parse_observable("MissingET0.MET")
\end{lstlisting}
\caption{Inherit from \texttt{Observable} and it will automatically retrieve the corresponding value if the physics object does have it.}
\label{code:MET}
\end{listing}

If you already have established a computation process and do not want to
consider the physical objects, we recommend referring to the second example
\ref{code:SumPt}. All you need to do is overwrite the \texttt{read} method,
defining how to compute the values of observables. The \texttt{events} is the
return value of \texttt{uproot.open}. You may need to adjust the calculation due
to the \texttt{events} and the underlying \texttt{awkward} array. It is
important to note that \texttt{\_value} must be an iterable object, such as a
list or array, to be correctly converted into an \texttt{awkward} array by
\texttt{hml}. Additionally, you should return \texttt{self} at the end, enabling
chain calls similar to other observables.

\begin{listing}[h]
\begin{lstlisting}
class SumPtOfJet0Muon0(Observable):
    def read(self, events):
        value = []
        pts1 = events["Jet.PT"].array()[:, 0:1]
        pts2 = events["Muon.PT"].array()[:, 0:1]

        # Loop over the events (slow)
        for pt1, pt2 in zip(pts1, pts2):
            value.append(sum(list(pt1) + list(pt2)))

        # Or manipulate the arrays directly (fast)
        pt1 = ak.pad_none(pts1, 1)[:, 0]
        pt2 = ak.pad_none(pts2, 1)[:, 0]
        pt1 = ak.fill_none(pt1, 0)
        pt2 = ak.fill_none(pt2, 0)

        self._value = value
        return self

register_observable(SumPtOfJet0Muon0, "mine")
obs = parse_observable("mine").read(events)
print(obs)
# mine: 1000 * float64
\end{lstlisting}
\caption{Overwrite the \texttt{read} method to define how to calculate the value of the observable.}
\label{code:SumPt}
\end{listing}

The third example \ref{code:strict_observable} changes the initialization. We
add constraints on physical objects, i.e. it could only be related to a single
physics object. There's also a new parameter for more flexibility. The
\texttt{read} part is the same as the second example. This is the strictest
observable but also the safest one.

\begin{listing}[h]
\begin{lstlisting}
class RestrictObservable(Observable):
    def __init__(self, new_parameter, physics_object, class_name=None):
        self.new_parameter = new_parameter
        super().__init__(physics_object, class_name, supported_objects = ["single"])

    def read(self, events):
        # Calculation of the observable
        return self
\end{lstlisting}
\caption{Define an observable with constraints on the type of physical objects and with a new parameter.}
\label{code:strict_observable}
\end{listing}

Currently, the naming convention is built upon the output of {\sc Delphes}
and does not support other formats yet. However, considering that different
analyses require data at different levels and in different formats, we plan to
gradually add support for other event formats in future versions, such as HEPMC,
LHE, etc.

\subsection{Representation} \label{representation}

To make high-energy physics data compatible with different analysis approaches,
it necessary to convert data into various representations. The review
\cite{Larkoski:2017jix} summarizes six representations of jets: ordered sets,
images, sequences, binary trees, graphs, and unordered sets. Built upon the
observable naming convention, we extend the representation to an event.
Currently, \texttt{hml} supports the (ordered) set and image representations. In
future versions, we will prioritize adding the graph representation and
corresponding neural networks.

The ordered set is one of the most commonly used representations. It arranges
physics-inspired observables in an arbitrary order to form a vector that
describes an event. The vectors from all events are then assembled into one
matrix by event, with the shape \texttt{(n\_events, n\_observables)}. Following
the naming convention, it is straightforward and concise to construct such a set
as illustrated in code example \ref{code:Set}.

\begin{listing}[h]
\begin{lstlisting}
from hml.representations import Set

r = Set(["jet0.pt", "muon0.charge", "fatjet0.mass"])
r.read(events)
# r.names
# ['jet0.pt', 'muon0.charge', 'fatjet0.mass']
# r.values
# [[189, None, 112],
#  [229, None, 65.6],
# ...,
# [251, None, 63.5],
# [311, None, 58.1]]
# -------------------------
# type: 1000 * 3 * ?float64
\end{lstlisting}
\caption{Use \texttt{Set} to represent the ordered set of observables.}
\label{code:Set}
\end{listing}

You need to package the observable names into a list and pass it into the
\texttt{Set}, then call the \texttt{read} method to get the values from events.
The values will be stored in the \texttt{values} attribute. Here, you can see
that we use \texttt{awkward} arrays to store data. For observables with the
correct physical object name but not existing (for example,
\texttt{muon0.charge}, when there are no muons produced in the event), we treat
its value as \texttt{None}. This way of handling missing values allows us to
follow the matrix operation habit: first build the data matrix, then deal with
the missing values.

For the image representation, we also use observable names to define the way of
fetching the data for its height, width, and channel, as shown in code example
\ref{code:init_Image}. The \texttt{read} method is used to read events as
before. Here, we construct an image of the leading fat jet, with all its
constituents' pseudorapidity, azimuthal angle, and transverse momentum.
Considering the significant amount of works that includes similar preprocessing
processes, we add them as the methods of the \texttt{Image} class. Since the
preprocessing often relates to sub-jets, it's necessary to add information about
sub-jets via the method \texttt{with\_subjets}: its parameters include the name
of constituents, clustering algorithm, radius, and the jet's minimum momentum.
The \texttt{translate} method moves the position of the leading sub-jet to the
origin, which reduces the complexity of position information helping speed up
the learning process. Next, for the sub-leading sub-jet, it can be rotated right
below the origin, making the features of the entire image more pronounced.
Lastly, the \texttt{pixelate} method is used to pixelate the data to make up an
real image. Since pixelation leads to a reduction in data precision, this step
is separated out, allowing for further studies on when to apply it and the
impacts of the order.

\begin{listing}[h]
\begin{lstlisting}
from hml.representations import Image

r = Image(
    height="fatjet0.constituents.phi",
    width="fatjet0.constituents.eta",
    channel="fatjet0.constituents.pt",
)
r.read(events)
r.with_subjets("fatjet0.constituents", "kt", 0.3, 0)
r.translate(origin="SubJet0")
r.rotate(axis="SubJet1", orientation=-90)
r.pixelate(size=(33, 33), range=[(-1.6, 1.6), (-1.6, 1.6)])
\end{lstlisting}
\caption{Use \texttt{Image} to represent a fat jet and preprocess it via sub-jets.}
\label{code:init_Image}
\end{listing}

\begin{figure}[h]
    \centering
    \includegraphics[width=0.22\textwidth]{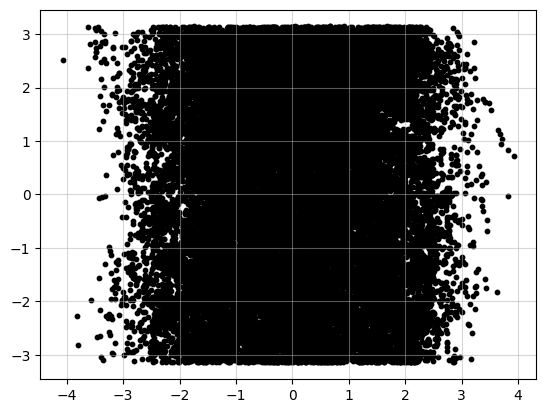}
    \includegraphics[width=0.22\textwidth]{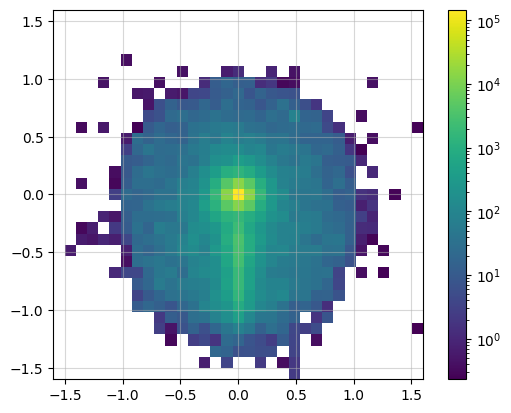}
\caption{The raw image and the pixelated image after preprocessing.}
\label{fig:Image}
\end{figure}

For convenience in displaying images, the \texttt{Image} class contains a
\texttt{show} method that can directly plot it as an image. Code example
\ref{code:show_Image} shows all the available parameters: the first two are used
to show the image as dots, and the last three parameters display a pixel-level
grid, enable the grid by default, and apply normalization over the whole image,
respectively. The figure \ref{fig:Image} shows an image representation before
and after preprocessing steps. For the final pixelated image,
\texttt{norm="log"} enhances its features more distinctly.

\begin{listing}[h]
\begin{lstlisting}
r.show(
    as_point=False,
    limits=None,
    show_pixels=False,
    grid=True,
    norm=None,
)
\end{lstlisting}
\caption{Use \texttt{show} to plot the image as a 2d heatmap if it has been pixelated or as a 2d scatter plot.}
\label{code:show_Image}
\end{listing}

After acquiring the original event data, it's time to filter it to obtain events
that satisfy specific criteria. In the old workflow, during the event loop, it
was common to manually include the calculation of observables and then apply
conditionals to filter events. We note that the \texttt(array) method in
\texttt{uproot} supports \texttt{cut} parameter. In \texttt{hml}, we utilize a
matrix-oriented programming style to change the filtering procedure into boolean
indexing; furthermore, we add logical operation syntax upon the observable
naming convention to make the definition of cuts intuitive, as shown in code
example \ref{code:cut}. \texttt{Cut} still has the similar \texttt{read} method.
The values form a one-dimensional boolean matrix, length of which is equal to
the number of events. It allows you to directly use it to filter other
observables via boolean indexing.

\begin{listing}[h]
\begin{lstlisting}
import uproot
from hml.approaches import Cut

tree = uproot.open("path/to/file.root")
events = tree["Delphes"]

cut = Cut("jet.size > 1 and fatjet.size > 0")
cut.read(events)
# cut.value
#[True,
# True,
# ...
# True,
# True]
# -----------------
# type: 1000 * bool

# Mask an observable
# obs.value[cut.value]
\end{lstlisting}
\caption{Cut values are boolean arrays that can be used to filter other observables.}
\label{code:cut}
\end{listing}

For the extend syntax of logical operations, i.e. how to combine multiple
conditions, we referr to the implementation of \texttt{uproot}: it uses the
bitwise logical operators of matrices, \texttt{\&} and \texttt{|} to replace
\texttt{and} or \texttt{or}, and adding parentheses to ensure priority. For
example, \texttt{(pt1 > 50) \& ((E1>100) | (E1<90))} expresses the condition
that \texttt{pt1} is greater than 50 and \texttt{E1} is either greater than 100
or less than 90. The expression is then parsed directly by Python: it is purely
matrix operation, without considering the case of {\sc Delphes} output. It
cannot handle such cuts: "all jets are required to have transverse momentum
greater than 10 GeV", whose data is of shape \texttt{n\_events, var}. It
undoubtedly requires users to rearrange the original data to make the dimensions
of the matrices consistent since the number of jets is not necessarily the same
among events. This essentially only filters values that meet the conditions, not
the events that meet the conditions. In conjunction with the observable naming
convention, we advocate simplifying the matrix logical operation syntax to
facilitate user input.

\begin{enumerate}
  \item Logical AND and logical OR are still represented by \texttt{and} or \texttt{or}. They are converted into bitwise logical operators \texttt{\&} and \texttt{|} automatically, avoiding too many parentheses;
  \item The result of the logical expression acts on the first dimension, that is, the dimension of events, to filter events;
  \item Involved observables must have the same dimensions to ensure the correctness of logical operations;
  \item At the end of a cut, default \texttt{[all]} represents a logical AND operation on all values of all observables in each event. It can be ignored and not written; \texttt{[any]} represents a logical OR operation on all values of all observables in each event. This syntax is suitable for cases where all of a certain observable or any one observable in the events needs to meet the conditions;
  \item Add support for \texttt{veto} at the beginning of a cut for cases where certain events need to be excluded.
\end{enumerate}

Below, we take specific literatures to demonstrate and explain the new syntax.
The original text will be presented first, followed by the corresponding cut in
the next line. In the literature \cite{Das:2017gke}:
\begin{enumerate}
  \item Muons $\mu^{ \pm}$ are identified with a minimum transverse momentum
  $p_T^\mu>10 \mathrm{GeV}$ and rapidity range $\left|\eta^\mu\right|<2.4$... \\

  \texttt{\small muon.pt >= 10 and -2.4 < muon.eta < 2.4} \\

  Here, we take all the muons and simplify the syntax for pseudorapidity within
  a certain range, which can be written consecutively. Here, the \texttt{and}
  represents the bitwise logical operator of the matrix. For each event, if
  there is no \texttt{[any]} at the end of the expression, it means that all
  values need to satisfy the condition;

  \item Only events with reconstructed di-muons having same sign are selected. \\

  \texttt{\small muon0.charge == muon1.charge} \\

  Here, we only need to judge whether the charges of the two muons are the same,
  without determining whether the number of muons is two. When there are fewer
  than two muons, the charge of one muon will be \texttt{None}, and such
  judgment will be automatically treated as \texttt{False};

  \item We identify the hardest fat-jet with the $W^{ \pm}$ candidate jet ($J$)
  and this is required to have $p_T^J>100 \mathrm{GeV}$. \\

  \texttt{\small fatjet0.pt >= 100}
\end{enumerate}

In the literature \cite{Bhardwaj:2018lma}:
\begin{enumerate}
  \item C3: we veto events if the OS di-muon invariant mass is less than 200 GeV. \\

  \texttt{\small muon0.charge != muon1.charge and muon0,muon1.inv\_mass > 200} \\

  \item C4: we apply a b-veto. \\

  \texttt{\small veto jet.b\_tag == 1 [any]} \\

  Use \texttt{veto} and \texttt{[any]} to indicate that for all jets, if any one
  of them is b-tagged, then the event is excluded.

  \item C5: we consider only events with a maximum MET of 60 GeV. \\

  \texttt{\small MissingET0.PT < 60} \\

  Using uppercase and lowercase is equivalent because there is only one missing
  energy, so it is represented by 0. The MET observable here actually also
  refers to the transverse momentum, so it can be represented by PT.

  \item C8: we choose events with $N$-subjettiness $\tau_{21}^{J_0}<0.4$. \\

  \texttt{\small fatjet0.tau21 < 0.4} \\

  The definition of this observable is provided by the {\sc Delphes} card. We
  have already defined its parsing method in the observables module, so here we
  can directly use its name.
\end{enumerate}

In the literature \cite{Chakraborty:2018khw}, the cuts of CBA-I for $M_N=600-900$ GeV:
\begin{enumerate}
    \item Jet-lepton separation $2.8<\Delta R(j, \ell)<3.8$ \\

    \texttt{\small 2.8 < jet0,lepton0.delta\_r < 3.8}
\end{enumerate}

In the literature \cite{Buonocore:2020erb}:
\begin{enumerate}
    \item The basic selections in our signal region require a lepton ($e$ or
    $\mu$) with $|\eta_{\ell}|<2.5$ and $p_{T, \ell}>500$ GeV ... \\

    \texttt{\small -2.5 < muon0.eta < 2.5 and muon0.pt > 500}. \\

    We take the muon as the example.

    \item ... veto events that contain additional leptons with
    $\left|\eta_{\ell}\right|<2.5$ and $p_{T, \ell}>7 \mathrm{GeV}$ \\

    \texttt{\small veto -2.5 < muon.eta < 2.5 and muon.pt > 7 [any]} \\

    In this cut, it's easier to use \texttt{veto} to exclude events with additional leptons.

    \item ... and impose a jet veto on subleading jets with
    $\left|\eta_j\right|<2.5$ and $p_{T, j}>30 \mathrm{GeV}$. \\

    \texttt{\small veto -2.5 < jet1.eta < 2.5 and jet1.pt > 30 [any]}
\end{enumerate}

In the literature \cite{Ngairangbam:2020ksz}:
\begin{enumerate}
    \item Photon-veto: Events having any photon with $p_T>15$ GeV in the central
    region, $|\eta|<2.5$ are discarded. \\

    \texttt{\small veto -2.5 < photon.eta < 2.5 and photon.pt > 15 [any]} \\

    \item $\tau$ and b-veto: No tau-tagged jets in $|\eta|<2.3$ with $p_T>18$
    GeV, and no b-tagged jets in $|\eta|<2.5$ with $p_T>20$ GeV are allowed. \\

    \texttt{\small veto jet.tau\_tag == 1 and -2.3 < jet.eta < 2.3 and jet.pt > 18 [any]} \\

    \item Alignment of MET with respect to jet directions: Azimuthal angle
    separation between the reconstructed jet with the MET to satisfy $\min
    (\Delta \phi(\mathbf{p}_T^{\mathrm{MET}}, \mathbf{p}_T^j))>0.5$ for up to
    four leading jets with $p_T>30 \mathrm{GeV}$ and $|\eta| < 4.7$. \\

    \texttt{\small jet:4,missinget0.min\_delta\_phi > 0.5 and jet:4.pt > 30 and -4.7 < jet:4.eta < 4.7} \\

    Users need to define the observable \texttt{MinDeltaPhi} in advance and
    register it with alias \texttt{min\_delta\_phi}.
\end{enumerate}

\subsection{Dataset} \label{dataset}

With the data representation and cuts defined previously, we can now proceed to
construct the dataset. Corresponding to data representations, we currently
offer two datasets: \texttt{SetDataset} and \texttt{ImageDataset}. Code example
\ref{code:init_SetDataset} shows the use of the dataset of an ordered set. Its
initialization requires names of the observables. Then still use the
\texttt{read} method to read events. For this \texttt{read}, we added two
additional parameters: \texttt{targets} is the integer label you assign to the
event, which is the target of convergence. Here we assign $1$ to denote events
as signals; \texttt{cuts} are your filtering criteria. Here we require the
number of jets to be more than $1$, and the number of leading fat jets to be
more than $0$. When you use multiple cuts, the result of each cut is applied to
the dataset one by one, which means they are connected by logical AND. When
splitting the dataset, you can use the \texttt{split} method. Its parameters are
the ratios for train, test, and validation sets, here we used 70\% of the data
as the training set, 20\% as the test set, and 10\% as the validation set.
Before saving the dataset, you can access the \texttt{samples} and
\texttt{targets} to view the stored data, which have already been converted into
\texttt{numpy} arrays. Finally, use the \texttt{save} method to save the dataset
to a zip compressed file with \texttt{.ds} suffix. Such a file can be loaded by
the \texttt{load\_dataset} function or \texttt{SetDataset.load} class method.

\begin{listing}[h]
\begin{lstlisting}
from hml.datasets import SetDataset, load_dataset

cuts = ["jet.size > 1 and fatjet.size > 0"]
set_ds = SetDataset(["jet0.pt", "jet1.pt", "fatjet0.mass"])

set_ds.read(events, targets=1, cuts=cuts)
set_ds.split(train=0.7, test=0.2, val=0.1)
set_ds.save("my_set_dataset.ds")
\end{lstlisting}
\caption{Use \texttt{SetDataset} to build a dataset representing each event as a set of observables.}
\label{code:init_SetDataset}
\end{listing}

To quickly view the distributions of the entire dataset, you can use the
\texttt{show} method. The code example \ref{code:show_SetDataset} shows all the
available parameters: \texttt{n\_feature\_per\_line} is the number of
observables to display per line, \texttt{n\_samples} is the number of events to
display, and \texttt{target} is the label of the events to display.

\begin{listing}[h]
\begin{lstlisting}[breaklines]{python}
set_ds.show(
    n_feature_per_line=3,
    n_samples=-1,
    target=None,
)
\end{lstlisting}
\caption{Use \texttt{show} to display the observable distributions of a set dataset.}
\label{code:show_SetDataset}
\end{listing}

The construction process of an image dataset is similar to that of an ordered
set, as shown in code example \ref{code:init_ImageDataset}. When initializing an
\texttt{Image}, you can directly configure the necessary preprocessing steps in
method chaining. When the dataset reads events afterwards, these steps will be
carried out in sequence.

\begin{listing}[h]
\begin{lstlisting}
from hml.representations import Image
from hml.datasets import ImageDataset

cuts = ["jet.size > 1 and fatjet.size > 0"]
image = (
    Image(
        height="fatjet0.constituents.phi",
        width="fatjet0.constituents.eta",
        channel="fatjet0.constituents.pt",
    )
    .with_subjets("fatjet0.constituents", "kt", 0.3, 0)
    .translate(origin="SubJet0")
    .rotate(axis="SubJet1", orientation=-90)
    .pixelate(size=(33, 33), range=[(-1.6, 1.6), (-1.6, 1.6)])
)
ds = ImageDataset(image)

ds.read(events, targets=1, cuts=cuts)
ds.split(train=0.7, test=0.2, val=0.1)
ds.save("my_dataset.ds")
\end{lstlisting}
\caption{Use \texttt{ImageDataset} to build a dataset representing each event as an image}
\label{code:init_ImageDataset}
\end{listing}

\begin{listing}[h]
\begin{lstlisting}
ds.show(
    limits=None,
    show_pixels=False,
    grid=True,
    norm=None,
    n_samples=-1,
    target=None,
)
\end{lstlisting}
\caption{Use \texttt{show} to plot the image of the dataset.}
\label{code:show_ImageDataset}
\end{listing}

The \texttt{show} method of an image dataset enable you to display events as an
image as shown in Example \ref{code:show_ImageDataset}. By default, the entire
dataset's images are compressed into one image. If the original dimensions are
\texttt{n\_events, height, width, channel}, the compressed dimensions will be
\texttt{height, width, channel}. While most parameters are the same as those in
the \texttt{show} method of \texttt{Image}, two additional parameters are added:
\texttt{n\_samples} is the number of events to display, and \texttt{target} is
the label of the events to display.

% Apply approaches ------------------------------------------------------------ %
\section{Apply approaches} \label{apply_approaches}

With well-prepared datasets, we can apply different approaches to identify rare
new physics signals. The \texttt{approaches} module includes cut-and-count,
trees, and neural networks. We will gradually add more in the coming versions.
The basic design principle of this module is minimal encapsulation to interface
with current frameworks, such as {\sc scikit-learn} \cite{sklearn_api}, {\sc TensorFlow},
and {\sc PyTorch}. Considering the simplicity, we adopt the Keras-style interface
design: decide approach structure at initialization, \texttt{compile} for
configuring the training process, \texttt{fit} for training the approach, and
\texttt{predict} for prediction on new data. {\sc Keras} is originally a
high-level encapsulation of {\sc TensorFlow}, but after the version 3, it begins to
support multiple backends, offering unprecedented flexibility, which is one of
the reasons we choose it. You should note that we only test the compatibility
with TensorFlow backend currently.

\subsection{CutAndCount} \label{cutandcount}

Cut-and-count (or cut-based analysis) is fundamental and widely used when
studying the impact of various observables on the final sensitivity. It provides
evidence to support the discovery of new particles and the verification of new
theories.

As the name suggests, it involves two steps: applying a series of cuts to
distinguish the signal from the background as much as possible, then counting
the number of events that pass the cuts. The subsequent distribution, such as
invariant mass, is used to determine the nature of the particles involved. These
cuts can be applied onto properties of specific particles, such as kinematic
quantities, charge, other observables, or other characteristics associated with
simulated collision events, like particle states in decay chains. Filtering the
data allows focusing on areas of interest, increasing the possibility of
discovering new physics.

Applying cuts involves some technique to make the final signal more evident.
Typically, one would plot the distribution of observables that reflect signal
characteristics and choose the area with a higher signal ratio as the cut range
based on manual judgment. There are two issues here: 1) manual judgment is
subjective and cannot guarantee the effect of the cut; 2) the observable
distributions that people observe is sometimes the source data without any cuts.
If there is an unavoidable association between observables, applying a cut will
affect the distribution of the next observable. Therefore, a more rigorous way
is to apply one cut first, plot the distribution of the next observable, then
determine the next cut, and so on.

\begin{listing}[h]
\begin{lstlisting}
from hml.approaches import CutAndCount

approach = CutAndCount(
    n_observables: int,
    n_bins: int = 50,
    topology="parallel", # or "sequential"
)
\end{lstlisting}
\caption{Initialize the CutAndCount approach}
\label{code:init_CutAndCount}
\end{listing}

Users can use \texttt{CutAndCount} to implement these two different strategies
of applying cuts. In code example \ref{code:init_CutAndCount}, we demonstrate
how to initialize a \texttt{CutAndCount} method. You need to specify the number
of involved observables, and then an internal \texttt{CutLayer} will be created
to automatically search for optimal cut values. For each observable, four
possible conditions are considered: the signal on the left side, right side,
middle, and both sides of the cut. Then, it calculates the user-specified loss
function for each case, choosing the one with minimum loss as the final cut.
\texttt{n\_bins} sets the granularity of the data when searching, i.e., the
number of bins for the distribution of each observable. A higher number of bins
can make the cut more precise but also increases the cost of calculation, which
also relates to the size of data and the complexity of its distribution. The
principle here is that as long as the data distributions become stable, then the
number of bins can be appropriately reduced without affecting the final result.
\texttt{topology} sets the order or the strategy that cuts are applied:
\texttt{parallel} means all cuts are independent, and the distributions referred
to come from the original data, while \texttt{sequential} considers the
correlation among cuts, with each cut applied on the basis of the previous one.

In code example \ref{code:compile_fit_CutAndCount}, we show how to configure and
train a \texttt{CutAndCount} approach. In the \texttt{compile} method, you can
specify the optimizer, loss function, and evaluation metrics. For
\texttt{CutAndCount}, the optimizer is unnecessary because it does not use
gradient descent methods internally, but rather finds the optimal cut values
through a search process. The loss function is used to evaluate the
effectiveness of each cut, while the evaluation metrics here will be used to
show the performance scores during the training, it is better to evaluate the
performance after training all at once. The \texttt{run\_eagerly=True} is
necessary. By default, {\sc Keras} uses a computational graph for
calculations, which is very efficient for training neural networks. However,
\texttt{CutAndCount} includes some custom calculations that are not yet fully
compatible within the computational graph, so it needs to be set to
\texttt{True}. In the \texttt{fit} method, you need to input the samples and
targets of the training set, where the batch size should be the minimum number
that can reflect the distribution pattern of the data. If your dataset is
relatively small and can fit entirely into the GPU's memory, you can set the
batch size to the size of the entire training set. Besides, the \texttt{epochs}
parameter is unnecessary, and the \texttt{callbacks} parameter has not been
implemented yet, but will be gradually added in future versions.

\begin{listing}[h]
\begin{lstlisting}
approach.compile(
    optimizer=None,
    loss="crossentropy",
    metrics=["accuracy"],
    run_eagerly=True,
)

approach.fit(
    dataset.train.samples,
    dataset.train.targets,
    batch_size=len(dataset.train.samples),
)
\end{lstlisting}
\caption{Configure and train the \texttt{CutAndCount} approach}
\label{code:compile_fit_CutAndCount}
\end{listing}

\subsection{Trees and Neural Networks} \label{trees_and_neural_networks}

Decision trees are a common method of classification, and there are many mature
frameworks available, such as {\sc TMVA} \cite{TMVA:2007ngy},
{\sc XGBoost} \cite{Chen:2016:XST:2939672.2939785}, {\sc scikit-learn},
etc. {\sc TensorFlow Decision Forests} \cite{GBBSP23} is also a good choice
as it also adopts the {\sc Keras} training style. Considering our preference
for the multi-backend support, we modify parts of the
\texttt{GradientBoostingClassifier} code from {\sc scikit-learn} to conform
to the same style.

Firstly, the original \texttt{fit} method is enhanced to handle input targets in
one-hot encoded format. Secondly, support for \texttt{Keras} metrics during the
training process, such as the commonly used \texttt{"accuracy"}. Thirdly, the
output of its \texttt{predict} method is changed to predict probabilities,
aligning it with \texttt{Keras}. Despite these changes, many parameters are
still not supported yet: many of the original initialization parameters are
related to early stopping, learning rate adjustments, etc., which in
\texttt{Keras} are implemented through callback functions. Moreover, loss
functions are not uniformly customizable in \texttt{scikit-learn}, so we do not
support changing it in \texttt{compile} method. In code example
\ref{code:use_GradientBoostedDecisionTree}, we demonstrate the basic usage.

\begin{listing}[h]
\begin{lstlisting}
from hml.approaches import GradientBoostedDecisionTree

bdt = GradientBoostedDecisionTree(...)
bdt.compile(
    # optimizer=None,
    # loss="crossentropy",
    metrics=["accuracy"],
)
bdt.fit(
    x_train,
    y_train,
    # callbacks
)
\end{lstlisting}
\caption{Basic usage of the modified \texttt{GradientBoostedDecisionTree} approach.}
\label{code:use_GradientBoostedDecisionTree}
\end{listing}

A starting point of \texttt{hml} is to provide researchers with existing deep
learning models so that they can conduct benchmark tests on their datasets and
select the optimal model. At this early development stage, we only offer two
basic models: \texttt{SimpleMLP} (multi-layer perceptron) and \texttt{SimpleCNN}
(convolutional neural network). In future versions, after thorough testing, we
will gradually add more existing models to provide the utmost convenience.

\begin{figure}[h]
  \centering
  \includegraphics[width=0.22\textwidth]{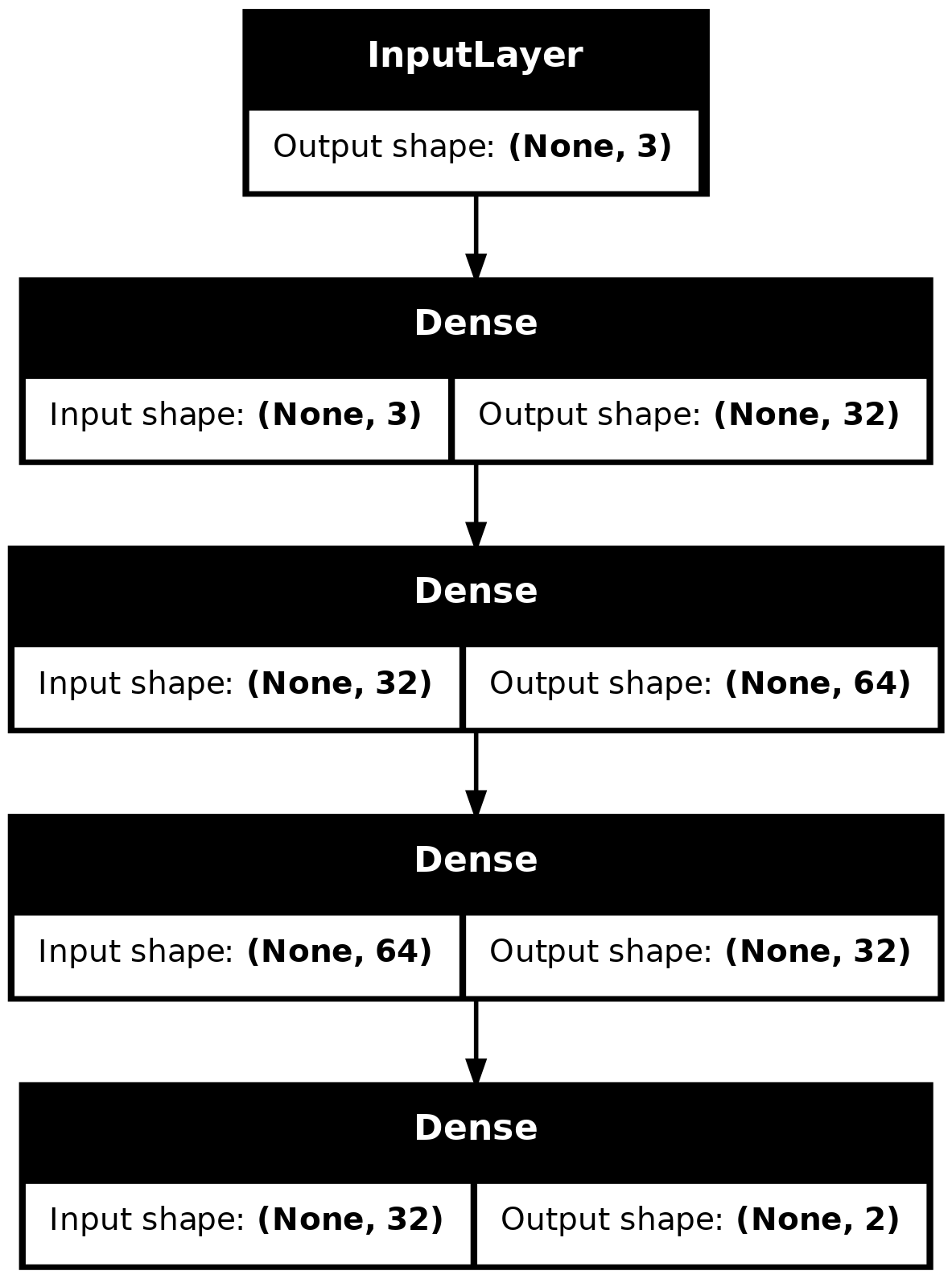}
  \includegraphics[width=0.22\textwidth]{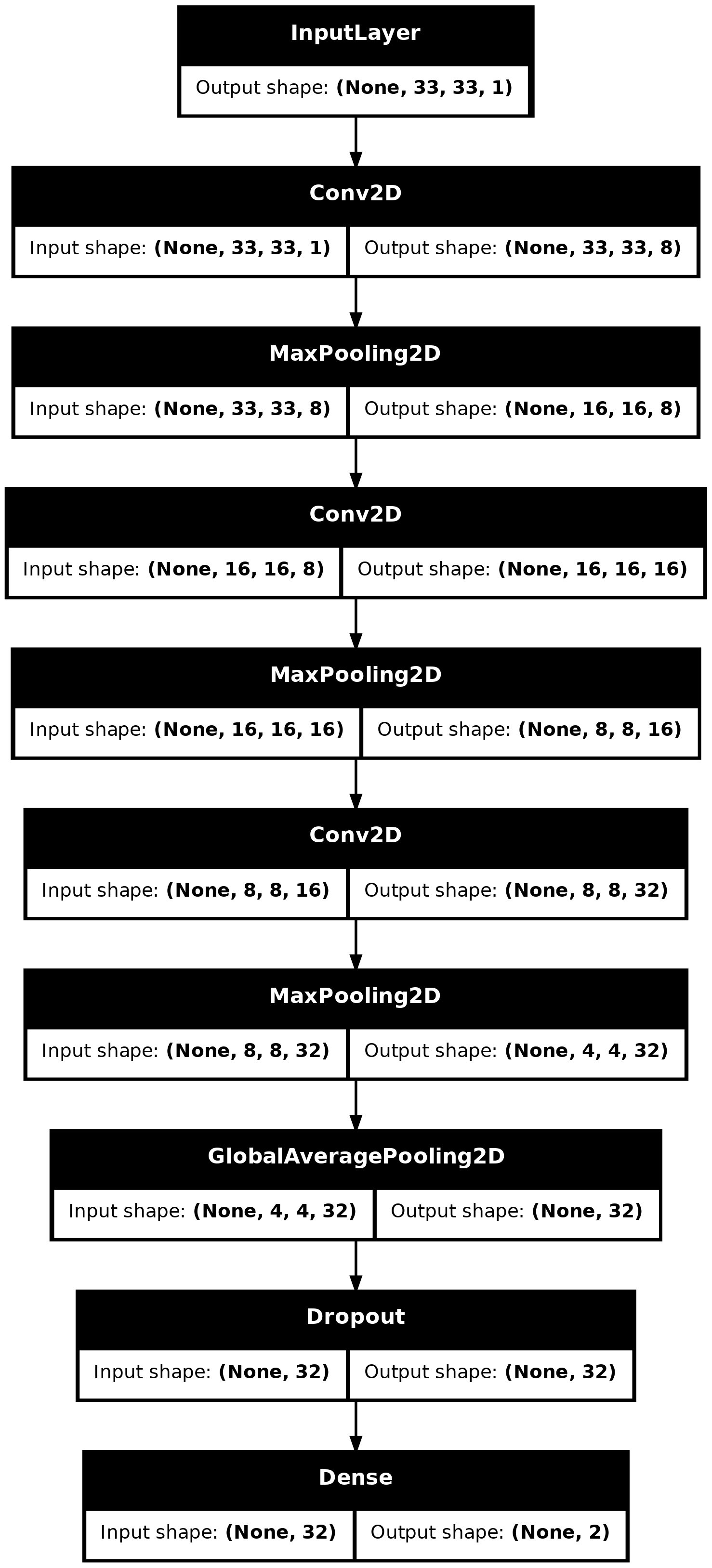}
  \caption{The structure of the \texttt{SimpleMLP} and \texttt{SimpleCNN} models.}
  \label{fig:structure}
\end{figure}

In {\sc Keras}, there are three ways to build a model: 1. Using
\texttt{Sequential} to stack layers, 2. Using the \textit{Functional} API to
construct more complex topologies, 3. Inheriting from \texttt{Model} to declare
a subclass for greater flexibility. Considering that the construction of many
models is quite complex and requires exposing a certain number of
hyper-parameters for tuning, we choose the third way to build the provided
models. The figure \ref{fig:structure} shows the structures of two models. The
\texttt{SimpleMLP} has $4386$ parameters with the inputs are three observables.
The \texttt{SimpleCNN} has $5960$ parameters with the inputs are images of shape
$(33, 33, 1)$. Both models are shallow and simple, so they do not consume too
much computing resource during the training and testing stages.

\subsection{Metrics} \label{metrics}

After training an approach on a dataset, it is often necessary to use various
metrics to assess its effectiveness. Unlike the classical accuracy score, which
is commonly used in classification tasks, in high-energy physics, the scarcity
of signals shifts the focus towards the signal significance, denoted by
$\sigma$. A higher value indicates a lower probability that the observed signal
is a result of background fluctuations alone. For instance, 3$\sigma$ is often
considered as an evidence of a signal, indicating that there is about a 0.27\%
chance of the signal being a statistical fluke. Meanwhile, 5$\sigma$ is the gold
standard in high-energy physics for claiming a discovery, corresponding to a
probability of roughly 1 in 3.5 million that the observed signal is due to
background noise. Equation \ref{equation:significance} is the formula for
calculating significance in \texttt{hml}. Note that here $S$ represents the
number of signals, and $B$ represents the number of backgrounds, which refer to
the number of simulated events when the cross section of the corresponding
process and integrated luminosity are not specified. In code example
\ref{code:use_MaxSignificance} we show how to use it.

\begin{equation}
\sigma = \frac{S}{\sqrt{S + B}}
\label{equation:significance}
\end{equation}

\begin{listing}[h]
\begin{lstlisting}
from hml.metrics import MaxSignificance

metric = MaxSignificance(
    cross_sections=[1, 1],
    luminosities=[1, 1],
    thresholds=None,
    class_id=1,
)

metric(y_true, y_test)
\end{lstlisting}
\caption{Use the \texttt{MaxSignificance} metric to evaluate performance of an trained approach.}
\label{code:use_MaxSignificance}
\end{listing}

For the built-in approaches, their outputs are probabilities for signal and
background. By default, only when the probability exceeds 0.5 do we consider it
as a signal or background. In \texttt{MaxSignificance}, we can change this
threshold by setting \texttt{thresholds}. By default, data with
\texttt{class\_id=1} is viewed as signal and 0 as background. Change it if the
targets for signal and background in your dataset are not like this.

In addition to significance, some literatures also include the background
rejection at a fixed signal efficiency as an evaluation metric, so we also
support it, shown in equation \ref{equation:background_rejection}. The higher
the background rejection rate, the fewer the number of background events that
are mistakenly classified as signals. In the example
\ref{code:use_RejectionAtEfficiency}, you can see the same way to use it. Note
that for both the metrics, if you call them multiple times, the values will be
averaged. So if you want to calculate them from scratch, you need to call the
\texttt{reset\_state} method.

\begin{equation}
\mathrm{rejection} = 1 /\left.\epsilon_b\right|_{\epsilon_s}
\label{equation:background_rejection}
\end{equation}

\begin{listing}[h]
\begin{lstlisting}
from hml.metrics import RejectionAtEfficiency

metric = RejectionAtEfficiency(
    efficiency,
    num_thresholds=200,
    class_id=None,
)

metric(y_true, y_test)
\end{lstlisting}
\caption{Use the \texttt{RejectionAtEfficiency} metric to evaluate the model's performance.}
\label{code:use_RejectionAtEfficiency}
\end{listing}

% Tagging --------------------------------------------------------------------- %
\section{Example: W boson tagging} \label{tagging}

To give users a complete understanding of the entire workflow, this section show
how to integrate various modules to complete a task of jet tagging. This example
serves merely as a proof of concept; users still need to conduct more
personalized analysis on this basis.

\subsection{Step 1: generate events}

We choose to simulate the production of highly boosted $W^{+}$ bosons that decay
into two jets, resulting in a single fat jet during event reconstruction. This
jet has distinct characteristics of mass and spatial distributions, making it
easier to identify using all built-in approaches.

In code example \ref{code:generate_signal}, we first import the event generator
module from \texttt{Madgraph5} using the \texttt{generate} method to create the
signal process, and then use the \texttt{output} method to save it to a
designated folder. To make $W^{+}$ bosons highly boosted, we leave the decay
chain unfinished here and constrain its $p_T$ range in code example
\ref{code:launch_signal}. If you want to view the output Feynman diagrams, after
the \texttt{output} command is completed, you can check them in the
\texttt{Diagrams} folder inside the output folder.

\begin{listing}[h]
\begin{lstlisting}
from hml.generators import Madgraph5

sig = Madgraph5(executable="mg5_aMC", verbose=0)
sig.generate("p p > w+ z")
sig.output("data/pp2wz")
\end{lstlisting}
\caption{Generate $W^{+}$ boson events using \texttt{Madgraph5}}
\label{code:generate_signal}
\end{listing}

\begin{figure}[h]
    \centering
    \includegraphics[width=0.4\textwidth]{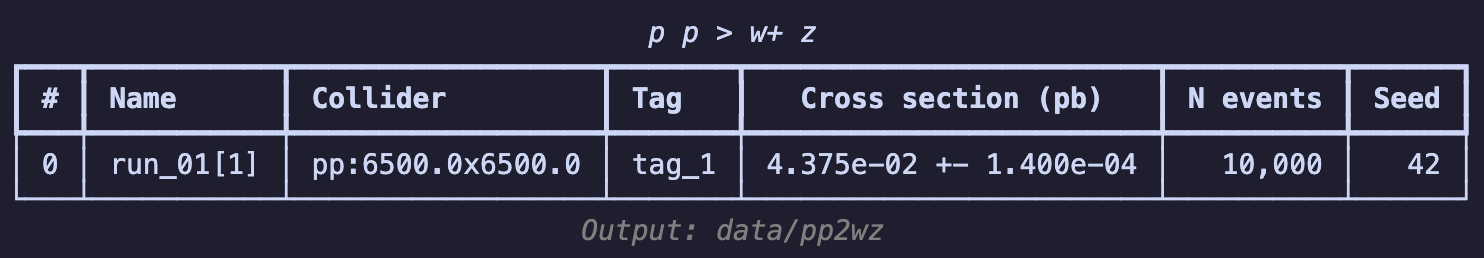}
    \includegraphics[width=0.4\textwidth]{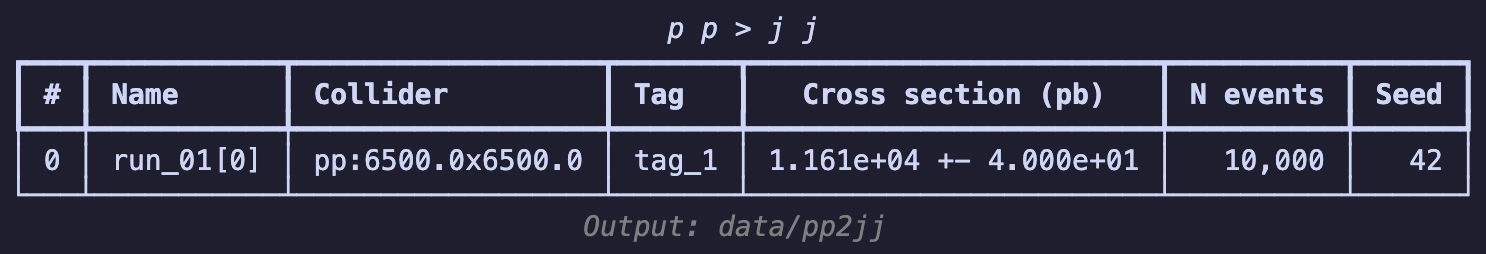}
    \caption{The summary table of the signal (upper) and background (lower) events.}
    \label{fig:summary_table}
\end{figure}

Then, use the \texttt{launch} method in code example \ref{code:launch_signal} to
start the simulation, turn on the \texttt{shower} and \texttt{detector}, and
turn off \texttt{madspin}. Following \cite{deOliveira:2015xxd}, set the \( p_T
\) range for the \( W^{+} \) boson from \( 250 \) to \( 350 \) GeV. When using
the default CMS delphes card with \( R=0.8 \) and the anti-\( k_T \) algorithm
to cluster the jets, you can obtain a fat jet that is expected to come from the
decay of the \( W^{+} \) boson. Use the \texttt{decay} method for further
specific decays. Set the random seed (\texttt{seed}) to 42 to ensure the
reproducibility of the results. When the simulation ends, use the
\texttt{summary} method to review the results (shown in figure
\ref{fig:summary_table}), akin to viewing results on the website of
\texttt{Madgraph5}.

\begin{listing}[h]
\begin{lstlisting}
sig.launch(
    shower="pythia8",
    detector="delphes",
    madspin="none",
    settings={
        "nevents": 10000,
        "pt_min_pdg": "{24: 250}",
        "pt_max_pdg": "{24: 300}",
    },
    decays=["w+ > j j", "z > vl vl~"],
    seed=42,
)

sig.summary()
\end{lstlisting}
\caption{Launch the simulation of $W^+$ boson events}
\label{code:launch_signal}
\end{listing}

The generation process for background events is similar
\ref{code:generate_background}, with the difference lying in the $p_T$ range
settings. Since the jets in this case do not originate from the decay of a
single particle, we directly restrict the $p_T$ range of the jets.

\begin{listing}[h]
\begin{lstlisting}
bkg = Madgraph5(executable="mg5_aMC", verbose=0)
bkg.generate("p p > j j")
bkg.output("data/pp2jj")
bkg.launch(
    shower="pythia8",
    detector="delphes",
    settings={
        "nevents": 10000,
        "ptj": 250,
        "ptjmax": 300,
    },
    seed=42,
)

bkg.summary()
\end{lstlisting}
\caption{Generate background events using \texttt{Madgraph5}}
\label{code:generate_background}
\end{listing}

After the event generation is complete, we start preparing the dataset. First,
in code example \ref{code:open_root_file}, we use \texttt{uproot} to open the
root file output by {\sc Delphes}, which stores branches categorized according
to physical objects. The \texttt{generators} module includes
\texttt{Madgraph5Run}, which conveniently retrieves information about the run,
such as the cross section and generated event files. Since it searches for files
produced in all sub-runs of a given run, even though the files for signal events
are stored in sub-runs, \texttt{run\_01} can also retrieve the corresponding
path correctly.

\begin{listing}[h]
\begin{lstlisting}
import uproot

from hml.generators import Madgraph5Run

sig_run = Madgraph5Run("./data/pp2wz", "run_01")
bkg_run = Madgraph5Run("./data/pp2jj", "run_01")

sig_events = uproot.open(sig_run.events()[0])
bkg_events = uproot.open(bkg_run.events()[0])
\end{lstlisting}
\caption{Use \texttt{uproot} to open the {\sc Delphes} output root file.}
\label{code:open_root_file}
\end{listing}

In code example \ref{code:prepare_set_dataset}, to avoid missing values of desired
observables, we use the previously mentioned extended logical operations to
apply cuts. For both types of datasets: \texttt{SetDataset} and
\texttt{ImageDataset}, it is required to have at least one fat jet and two
regular jets. The \texttt{read} method supports entering multiple cuts, thus
there is no need to use the \texttt{Cut} class to parse the expressions first.
When reading events, integer labels are assigned separately for signal and
background events. Before saving locally, the data is split into a training set
and a testing set at a 7:3 ratio.

\begin{listing}[h]
\begin{lstlisting}
from hml.datasets import SetDataset

cut = "fatjet.size > 0 and jet.size > 1"
set_ds = SetDataset(
    [
        "fatjet0.mass",
        "fatjet0.tau21",
        "jet0,jet1.delta_r",
    ]
)
set_ds.read(sig_events, 1, [cut])
set_ds.read(bkg_events, 0, [cut])

set_ds.split(0.7, 0.3)
set_ds.save("./data/wjj_vs_qcd_set.ds")
\end{lstlisting}
\caption{Prepare the set dataset.}
\label{code:prepare_set_dataset}
\end{listing}

For the image dataset, it is first to construct the representation of the data:
namely, what observables should constitute the images and what preprocessing
steps should be taken. In code example
\ref{code:construct_image_representation}, $\phi$, $\eta$, and $p_T$ of all
constituents from the leading fat jet are used as the data source for height,
width, and channel of an image. Before preprocessing, \texttt{with\_subjets} is
used to recluster constituents to add information about the subjets. Since the
distance between the two sub-jets will not be less than $0.3$ according the
previous equation, it is safe to use the $k_T$ algorithm with $R=0.3$. Then,
\texttt{translate} and \texttt{rotate} are used to translate and rotate the
image, aligning the information of the two sub-jets. Finally, \texttt{pixelate}
is used to pixelate the image; the size here is $33 \times 33$, with a range of
$(-1.6, 1.6)$, and an equivalent precision of around $0.1$. This precision does
not match the precision in the detector card. For simplicity, we take this fixed
precision. In code example \ref{code:prepare_image_dataset}, we show how to
prepare the image dataset.

\begin{listing}[h]
\begin{lstlisting}
from hml.representations import Image
from hml.datasets import ImageDataset

image = (
    Image(
        height="fatjet0.constituents.phi",
        width="fatjet0.constituents.eta",
        channel="fatjet0.constituents.pt",
    )
    .with_subjets(
        constituents="fatjet0.constituents",
        algorithm="kt",
        r=0.3,
        min_pt=0,
    )
    .translate(origin="SubJet0")
    .rotate(axis="SubJet1", orientation=-90)
    .pixelate(
        size=(33, 33),
        range=[(-1.6, 1.6), (-1.6, 1.6)],
    )
)
\end{lstlisting}
\caption{Construct the representation of the image dataset.}
\label{code:construct_image_representation}
\end{listing}

\begin{listing}[h]
\begin{lstlisting}
image_ds = ImageDataset(image)
image_ds.read(sig_events, 1, [cut])
image_ds.read(bkg_events, 0, [cut])

image_ds.split(0.7, 0.3)
image_ds.save("./data/wjj_vs_qcd_image.ds")
\end{lstlisting}
\caption{Prepare the image dataset.}
\label{code:prepare_image_dataset}
\end{listing}

\begin{figure}[h]
    \centering
    \includegraphics[width=0.46\textwidth]{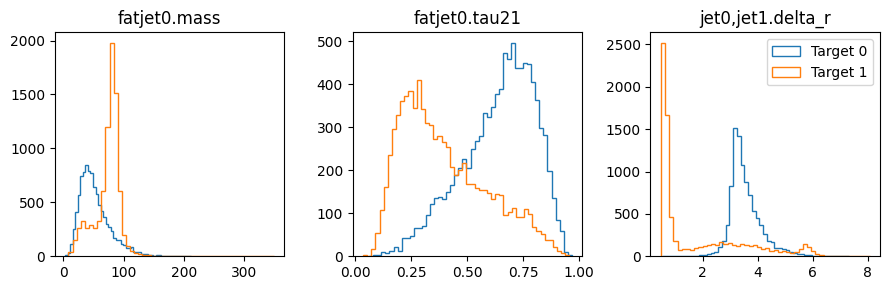}
    \caption{Feature distributions of the set dataset.}
    \label{fig:set_dataset}
\end{figure}

\begin{figure}[h]
    \centering
    \includegraphics[width=0.22\textwidth]{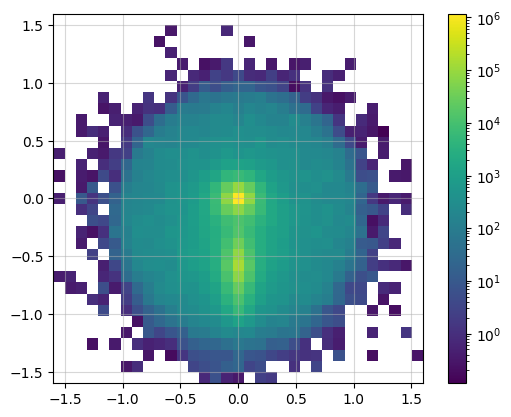}
    \includegraphics[width=0.22\textwidth]{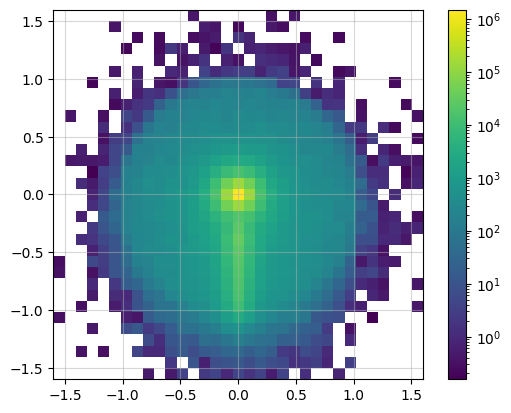}
    \caption{Combined images of signal (left) and background (right) events.}
    \label{fig:image_dataset}
\end{figure}

After constructing the datasets, we count signal and background samples in code
example \ref{code:count_show_datasets} to avoid introducing artificial bias. We
then use the dataset's \texttt{show} method to display the distribution of
observables, as illustrated in figure \ref{fig:set_dataset} and figure
\ref{fig:image_dataset}.

\begin{listing}[h]
\begin{lstlisting}
import numpy as np

print(np.unique(set_ds.targets, return_counts=True))
# (array([0, 1], dtype=int32), array([9782, 8281]))

set_ds.show()
image_ds.show(norm="log", target=0, show_pixels=True)
image_ds.show(norm="log", target=1, show_pixels=True)
\end{lstlisting}
\caption{Count the number of signals and backgrounds and show the distribution of the datasets.}
\label{code:count_show_datasets}
\end{listing}

It is observed that the number of signal and background instances is
approximately equal, and the observables used clearly reflect the distinct
characteristics of each, which is beneficial for our subsequent classification
tasks. In the display of the image dataset, we show a merged image of the signal
and background. It can be seen that the signal's fat jet image prominently
features two sub-jets, whereas the sub-jets in the background are less distinct.

\subsection{Step 3: apply approaches}

Once the datasets are prepared, let us import all available approaches for
training. These approaches learn the differences between the signal and
background. Subsequently, we use a dictionary and the built-in metrics to gather
their performance, which will then be presented as a benchmark test. First,
import all the necessary packages as \ref{code:import_packages}, which we have
roughly categorized to facilitate understanding of their purposes.

\begin{listing}[h]
\begin{lstlisting}
# General
import numpy as np
import matplotlib.pyplot as plt
from keras import ops
from rich.table import Table

# Dataset
from sklearn.preprocessing import MinMaxScaler
from hml.datasets import load_dataset

# Approaches
from hml.approaches import CutAndCount as CNC
from hml.approaches import GradientBoostedDecisionTree as BDT
from hml.approaches import SimpleCNN as CNN
from hml.approaches import SimpleMLP as MLP

# Evaluation
from keras.metrics import Accuracy, AUC
from sklearn.metrics import roc_curve
from hml.metrics import MaxSignificance, RejectionAtEfficiency

# Save and load
from hml.approaches import load_approach
\end{lstlisting}
\caption{Import necessary packages for the benchmark test.}
\label{code:import_packages}
\end{listing}

The selected evaluation metrics are accuracy, AUC (Area Under the Curve), signal
significance, and background rejection rate at fixed signal efficiency. These
metrics are commonly used in high-energy physics and help us better understand
the approaches' performance. In code example \ref{code:get_result}, we define a
function to retrieve the evaluation metrics of an approach.

\begin{listing}[h]
\begin{lstlisting}
results = {}

def get_result(approach, x_test, y_test):
    y_pred = approach.predict(x_test, verbose=0)
    fpr, tpr, _ = roc_curve(y_test, y_pred[:, 1])
    result = {
        approach.name: {
            "acc": Accuracy()(y_test, y_pred.argmax(axis=1)),
            "auc": AUC()(y_test, y_pred[:, 1]),
            "sig": MaxSignificance()(y_test, y_pred),
            "r50": RejectionAtEfficiency(0.5)(y_test, y_pred),
            "r99": RejectionAtEfficiency(0.99)(y_test, y_pred),
            "fpr": fpr,
            "tpr": tpr,
        }
    }

    return result
\end{lstlisting}
\caption{Define a function to get the evaluation metrics of a model.}
\label{code:get_result}
\end{listing}

For approaches, cut-and-count and the decision tree, they are not sensitive to
the scale of features, so we can directly import the dataset (code example
\ref{code:load_set_dataset}) and start training (code example
\ref{code:train_cut_and_count}). Two different topologies of the cut-and-count
are taken to demonstrate how the order of applying cuts impacts performance.

\begin{listing}[h]
\begin{lstlisting}
ds = load_dataset("./data/wjj_vs_qcd_set.ds")

x_train, y_train = ds.train.samples, ds.train.targets
x_test, y_test = ds.test.samples, ds.test.targets
\end{lstlisting}
\caption{Load the set dataset for training cut-and-count and decision tree approaches.}
\label{code:load_set_dataset}
\end{listing}

\begin{listing}[h]
\begin{lstlisting}
cnc1 = CNC(
    n_observables=3,
    topology="parallel",
    name="cnc_parallel",
)
cnc1.compile(
    optimizer="adam",
    loss="crossentropy",
    metrics=["accuracy"],
    run_eagerly=True,
)
cnc1.fit(x_train, y_train, batch_size=len(x_train))

cnc2 = CNC(
    n_observables=3,
    topology="sequential",
    name="cnc_sequential",
)
cnc2.compile(
    optimizer="adam",
    loss="crossentropy",
    metrics=["accuracy"],
    run_eagerly=True,
)
cnc2.fit(x_train, y_train, batch_size=len(x_train))

results.update(get_result(cnc1, x_test, y_test))
results.update(get_result(cnc2, x_test, y_test))
\end{lstlisting}
\caption{Train the cut-and-count approaches with two different topologies.}
\label{code:train_cut_and_count}
\end{listing}

\begin{listing}[h]
\begin{lstlisting}
bdt = BDT(name="bdt")
bdt.compile(metrics=["accuracy"])
bdt.fit(x_train, y_train)

results.update(get_result(bdt, x_test, y_test))
\end{lstlisting}
\caption{Train the decision tree approach.}
\label{code:train_decision_tree}
\end{listing}

The input for the multilayer perceptron requires the use of
\texttt{MinMaxScaler} to scale the features within the 0-1 range, as detailed in
code example \ref{code:load_set_dataset_mlp}, which aids in the model's rapid
convergence. We train the model for 100 epochs with a batch size of 128. In code
example \ref{code:train_mlp}, we illustrate the training process.

\begin{listing}[h]
\begin{lstlisting}
ds = load_dataset("./data/wjj_vs_qcd_set.ds")

x_train, y_train = ds.train.samples, ds.train.targets
x_test, y_test = ds.test.samples, ds.test.targets

scaler = MinMaxScaler()
x_train = scaler.fit_transform(x_train)
x_test = scaler.transform(x_test)
\end{lstlisting}
\caption{Load the set dataset for training the multilayer perceptron approach.}
\label{code:load_set_dataset_mlp}
\end{listing}

\begin{listing}[h]
\begin{lstlisting}
mlp = MLP(name="mlp", input_shape=x_train.shape[1:])
mlp.compile(
    optimizer="adam",
    loss="crossentropy",
    metrics=["accuracy"],
)
mlp.fit(
    x_train,
    y_train,
    batch_size=128,
    epochs=100,
)
results.update(get_result(mlp, x_test, y_test))
\end{lstlisting}
\caption{Train the multilayer perceptron approach.}
\label{code:train_mlp}
\end{listing}

For the convolutional neural network, we employ two different preprocessing
methods. One scales each image by its maximum value, while the other applies a
logarithmic transformation to each pixel value. Given the large variations in
pixel intensity in jet images, scaling directly by the maximum value might
result in excessively small pixel values, whereas logarithmic transformation
better preserves information. In code example \ref{code:load_image_dataset} and
\ref{code:normalize_pixel_values}, we load the image dataset and demonstrate
these two distinct preprocessing techniques.

\begin{listing}[h]
\begin{lstlisting}
ds = load_dataset("./data/wjj_vs_qcd_image.ds")

x_train, y_train = ds.train.samples, ds.train.targets
x_test, y_test = ds.test.samples, ds.test.targets

non_zero_train = x_train.reshape(x_train.shape[0], -1).sum(1) != 0
non_zero_test = x_test.reshape(x_test.shape[0], -1).sum(1) != 0

x_train, y_train = x_train[non_zero_train], y_train[non_zero_train]
x_test, y_test = x_test[non_zero_test], y_test[non_zero_test]

x_train = (
    x_train.reshape(len(x_train), -1)
    / x_train.reshape(len(x_train), -1).max(1, keepdims=True)
).reshape(x_train.shape)
x_test = (
    x_test.reshape(len(x_test), -1)
    / x_test.reshape(len(x_test), -1).max(1, keepdims=True)
).reshape(x_test.shape)
x_train = x_train[..., None]
x_test = x_test[..., None]
\end{lstlisting}
\caption{Load the image dataset and normalize the pixel values with the maximum value.}
\label{code:load_image_dataset}
\end{listing}

\begin{listing}[h]
\begin{lstlisting}
cnn1 = CNN(name="cnn_max", input_shape=x_train.shape[1:])
cnn1.compile(
    optimizer="adam",
    loss="crossentropy",
    metrics=["accuracy"],
)
cnn1.fit(
    x_train,
    y_train,
    epochs=100,
    batch_size=128,
)
results.update(get_result(cnn1, x_test, y_test))
\end{lstlisting}
\caption{Train the convolutional neural network approach with the maximum value normalization.}
\label{code:train_cnn_max}
\end{listing}

\begin{listing}[h]
\begin{lstlisting}
...
x_train = np.log(x_train + 1)
x_test = np.log(x_test + 1)
...

cnn2 = CNN(name="cnn_log", input_shape=x_train.shape[1:])
...
\end{lstlisting}
\caption{Normalize the pixel values by taking the logarithm. \texttt{"..."} indicates the same code as in \ref{code:load_image_dataset} and \ref{code:train_cnn_max}.}
\label{code:normalize_pixel_values}
\end{listing}

Finally, we present a comparison of performance using code example
\ref{code:compare_performance}. The results are shown in table
\ref{tab:approach_comparison}.

From the significance column, we observe that for the cut-and-count method, the
sequential topology, which considers the impacts among cuts, performs better
than the parallel topology. For convolutional neural networks, the performance
of logarithmic scaling is roughly equivalent to that using maximum value
scaling. The multilayer perceptron and decision trees, which utilize features
with clear distinctions, exhibit the best performance. For more practical
problems, we can apply different approaches to the dataset and then select the
most suitable one based on various performance metrics.

\begin{listing}[h]
\begin{lstlisting}
table = Table(
    "Name", "ACC", "AUC", "Significance", "R50", "R99", title="Approach Comparison"
)

for name, metrics in results.items():
    table.add_row(
        name,
        f"{metrics['acc']:.6f}",
        f"{metrics['auc']:.6f}",
        f"{metrics['sig']:.6f}",
        f"{metrics['r50']:.6f}",
        f"{metrics['r99']:.6f}",
    )

print(table)
\end{lstlisting}
\caption{Compare the performance of different approaches.}
\label{code:compare_performance}
\end{listing}

\begin{table*}[t]
    \centering
    \begin{tabular}{|l|r|r|r|r|r|}
    \hline
    \textbf{Name} & \textbf{ACC} & \textbf{AUC} & \textbf{Significance} & \textbf{R50} & \textbf{R99} \\ \hline
    \texttt{cnc\_parallel} & 0.750323 & 0.728121 & 33.660892 & 4.005174 & 1.000000 \\ \hline
    \texttt{cnc\_sequential} & 0.787784 & 0.769440 & 36.557026 & 4.712174 & 1.000000 \\ \hline
    \texttt{bdt} & 0.902011 & 0.955063 & 44.368549 & 117.804291 & 2.146139 \\ \hline
    \texttt{mlp} & 0.900904 & 0.956274 & 44.205276 & 117.804291 & 2.124265 \\ \hline
    \texttt{cnn\_max} & 0.806827 & 0.867769 & 38.444225 & 17.089737 & 1.188322 \\ \hline
    \texttt{cnn\_log} & 0.809452 & 0.876692 & 38.732323 & 19.042860 & 1.276514 \\ \hline
    \end{tabular}
\caption{Comparison of different approaches}
\label{tab:approach_comparison}
\end{table*}

% Summary --------------------------------------------------------------------- %
\section{Summary} \label{summary}

In the current era where machine learning models are rapidly evolving, it is
worthwhile to explore how to use them more conveniently in high-energy physics
for searching new physical signals. In this paper, we introduce the \texttt{hml}
Python package, which offers a streamlined workflow from event generation to
performance evaluation. The simplified process and control over random seeds
significantly enhance the reproducibility of the final analysis results.

We propose a naming convention for observables, which enables users to easily
extract the required data from events output by {\sc Delphes}. Additionally,
we extend the cut expression syntax originally in {\sc uproot} to make it
more user-friendly and compatible with {\sc Delphes} output formats. This
convention is also utilized in our dataset construction process, helping users
to quickly and conveniently build datasets. Based on this naming convention, we
implement a transformation from the output of event generators to datasets
usable by various analysis approaches. Moreover, the \texttt{show} method
included in datasets enables users to display data either as 1D distributions or
2D images, facilitating the adjustment of observable selections based on
observed differences.

We have adopted the interface style of {\sc Keras} to standardize traditional
methods such as the cut-and-count technique and decision trees. Furthermore, the
cut-and-count approach supports automatic searching for the optimal cut
positions, significantly reducing the workload for users. Additionally, we have
incorporated commonly used evaluation metrics in high-energy physics, such as
signal significance and background rejection rate at fixed signal efficiency.
These metrics help users better understand the performance of the models.

We demonstrate the complete workflow through a practical example. It intuitively
showcases the usage of \texttt{hml}. Currently, \texttt{hml} is continuously
being updated. We plan to incorporate more existing deep learning models,
datasets, and extend to graph representations of data to further enhance its
capabilities.

\section*{Acknowledgments}
H.S. is supported by the National Natural Science Foundation of China under
Grants No.~12075043.

\bibliographystyle{spphys}
\bibliography{references}

\end{document}